\documentclass[12pt]{article}
\usepackage{graphicx}
\usepackage{epsf} 
\def\beq{\begin{equation}}
\def\eeq{\end{equation}}
\def\bea{\begin{eqnarray}}
\def\eea{\end{eqnarray}}
\def\nn{\nonumber}
\def\sss{\scriptscriptstyle}
\def\roughly#1{\mathrel{\raise.3ex\hbox
{$#1$\kern-.75em\lower1ex\hbox{$\sim$}}}}
\def\lsim{\roughly<}

\def\bd{B_d^0}

\def\btos{{\bar b} \to {\bar s}}
\def\bsss{{\bar b} \to {\bar s} s {\bar s}}
\def\ks{K_{\sss S}}

\def\bra#1{\left\langle #1\right|}
\def\ket#1{\left| #1\right\rangle}

\def\BKstar{B\to\phi K^*}
\def\BKs{\bd\to\phi\ks}
\def\fT{f_{\sss T}}
\def\fL{f_{\sss L}}
\def\fTfL{f_{\sss T}/f_{\sss L}}

\begin{document}

\begin{flushright}
UMiss-HEP-2008-06 \\
UdeM-GPP-TH-08-172 \\
\end{flushright}

\begin{center}
\bigskip
{\Large \bf \boldmath $\BKstar$, $\BKs$ and New Physics} \\
\bigskip
\bigskip
Alakabha Datta $^{a,}$\footnote{datta@phy.olemiss.edu},
Maxime Imbeault
$^{b,}$\footnote{maxime.imbeault@umontreal.ca},
and David London $^{b,}$\footnote{london@lps.umontreal.ca}
\end{center}
\begin{flushleft}
~~~~~~~~~~~$a$: {\it Dept of Physics and Astronomy, 108 Lewis
Hall,}\\
~~~~~~~~~~~~~~~{\it University of Mississippi, Oxford, MS
38677-1848, USA}\\
~~~~~~~~~~~$b$: {\it Physique des Particules, Universit\'e de
Montr\'eal,}\\
~~~~~~~~~~~~~~~{\it C.P. 6128, succ. centre-ville,
Montr\'eal, QC, Canada H3C 3J7}\\
\end{flushleft}

\begin{center}
\bigskip (\today)
\vskip0.5cm {\Large Abstract\\} \vskip3truemm
\parbox[t]{\textwidth}{We consider the new-physics (NP)
solution to the polarization puzzle in $\BKstar$ decays.  We
note that any such solution must reproduce the data in
$\BKs$, where there are disagreements with the standard model
in CP-asymmetry measurements.  We examine 10 NP operators, of
$S/P$, $V/A$ and $T$ variety. We find that, as long as $\BKs$
exhibits large CP-violating effects, no single operator can
explain the observations in both $\BKstar$ and $\BKs$.  For
2-NP-operator solutions, there are four possibilities, all of
$S/P$ type, which are presently allowed.  We discuss ways of
distinguishing among these solutions in the future. Models
which contain only $V/A$ operators, such as those with
supersymmetry or extra $Z'$ bosons, cannot explain both
$\BKs$ and $\BKstar$ data. On the other hand, the
two-Higgs-doublet model, which has only $S/P$ operators, is
favored.}
\end{center}

\thispagestyle{empty}
\newpage
\setcounter{page}{1}
\baselineskip=14pt

$B\to V_1V_2$ decays (the $V_i$ are vector mesons) really
represent three transitions when the spins of the $V_i$ are
taken into account.  That is, the final state can be
transversely (2 states) or longitudinally (1 state)
polarized.  A naive calculation within the standard model
(SM) shows that the transverse amplitudes are suppressed by a
factor of size $m_{\sss V}/m_{\sss B}$ ($V$ is one of the
vector mesons) with respect to the longitudinal
amplitude. One then expects the fraction of transverse
decays, $\fT$, to be much less than the fraction of
longitudinal decays, $\fL$.

However, it was observed that these two fractions are roughly
equal in the decay $\BKstar$: $\fTfL \simeq 1$
\cite{phiK*,phiK0*recent,phiK+*recent}.  This is the
``polarization puzzle.'' If one goes beyond the naive SM,
there are two explanations \cite{DGLNS} which account for
this surprising result: penguin annihilation (PA)
\cite{Kagan} and non-perturbative rescattering \cite{soni,
SCET}.  Still, there are question marks associated with both
of these.  First, PA is a subleading amplitude that is power
suppressed by $O(1/m_b)$.  Second, for rescattering, it is
not obvious whether such a non-perturbative effect is of
leading or subleading order. Hence, the SM explanations of
the large $\fTfL$ generally require enhanced subleading
amplitudes (certain for PA; possible for rescattering).

One can also explain the $\fTfL$ measurement by introducing
physics beyond the SM.  Suppose there are new-physics (NP)
contributions to the $\bsss$ quark-level amplitude. If their
form is chosen correctly, these will contribute dominantly to
$\fT$ in $\BKstar$ and not to $\fL$, so that one can
reproduce the measured value of $\fTfL$ if the NP amplitude
has the right size \cite{BphiK*NP}. In this paper, we explore
the NP explanation -- we assume that neither PA nor
non-perturbative rescattering produce dominant contributions
to the transverse amplitudes, and are therefore not the
explanation of the measurement of a large $\fTfL$.

\begin{table}[tbh]
\center
\begin{tabular}{|c|c|}
\hline
BR & $8.3^{+1.2}_{-1.0} \times 10^{-6}$ \\
$S_{\sss CP}$ & $0.44^{+0.17}_{-0.18}$ \\
$A_{\sss CP}$ & $0.23\pm 0.15$ \\
\hline
\end{tabular}
\caption{Measurements of $\BKs$.  Included are the branching
  fraction (BR) \cite{pdg,BR,HFAG}, the indirect
  (mixing-induced) CP asymmetry ($S_{\sss CP}$)
  \cite{HFAG,CPasym}, and the direct CP symmetry ($A_{\sss
  CP}$) \cite{HFAG,CPasym}.}
\label{tab:dataPV}
\end{table}

Now, any NP contribution to $\bsss$ will also affect $\BKs$.
Thus, any constraints on such NP must take into account the
measurements of both $\BKstar$ and $\BKs$.  The $\BKs$ data
are shown in Table 1. The CP-violating observables are
particularly intriguing.  Within the SM, including small
corrections, the indirect (mixing-induced) CP asymmetry in
$\BKs$ ($S_{\sss CP}$) is expected to be a bit larger than
that in charmonium $\bd$ decays \cite{chua}, found to be
$S_{\sss CP}({\rm charmonium}) = 0.672 \pm 0.024$
\cite{HFAG}. In addition, the direct CP asymmetry in $\BKs$
($A_{\sss CP}$) is expected to vanish.  In other words, the
central values of both of these measurements exhibit
disagreements with the expectations of the SM.  This provides
a hint of NP in $\btos$ transitions\footnote{A more
significant hint (signal?) of NP, also in $\btos$
transitions, is provided by $B\to\pi K$ decays, where the
disagreement with the SM has reached the $5\sigma$ level
\cite{HFAG}, assuming that $|C/T|$ is small, as is expected
in the SM \cite{BL} ($C$ and $T$ are diagrams contributing to
the decay \cite{GHLR}).}.

On the other hand, the errors are sufficiently large that the
discrepancies are only at the level of $\lsim 2\sigma$.  This
means that the hint of NP is not statistically significant.
It also means that any constraints on NP in $\BKstar$ are not
that strong.  For this reason, in this paper we also perform
the analysis in the scenario in which the hint of NP in
$\BKs$ becomes a true signal in the future.  That is, in this
case we assume that the future measurements of $S_{\sss CP}$
and $A_{\sss CP}$ stay at their present central values, but
the errors are reduced by a factor of 2. If the discrepancies
with the SM in $\BKs$ get more pronounced, it will be
necessary to consider the results from this second scenario.

We assume that there is a NP contribution to $\bsss$.  We
consider operators of the form
\bea
O_{\sss LL}^{\sss V/A} &=& \bar s \gamma_\mu (1-\gamma_5) b
\, \bar s \gamma^\mu (1-\gamma_5) s~, \nn\\
O_{\sss LR}^{\sss V/A} &=& \bar s \gamma_\mu (1-\gamma_5) b
\, \bar s \gamma^\mu (1+\gamma_5) s~, \nn\\
O_{\sss RL}^{\sss V/A} &=& \bar s \gamma_\mu (1+\gamma_5) b
\, \bar s \gamma^\mu (1-\gamma_5) s~, \nn\\
O_{\sss RR}^{\sss V/A} &=& \bar s \gamma_\mu (1+\gamma_5) b
\, \bar s \gamma^\mu (1+\gamma_5) s~, \nn\\
O_{\sss LL}^{\sss S/P} &=& \bar s (1-\gamma_5) b \, \bar s
(1-\gamma_5) s~, \nn\\
O_{\sss LR}^{\sss S/P} &=& \bar s (1-\gamma_5) b \, \bar s
(1+\gamma_5) s~, \nn\\
O_{\sss RL}^{\sss S/P} &=& \bar s (1+\gamma_5) b \, \bar s
(1-\gamma_5) s~, \nn\\
O_{\sss RR}^{\sss S/P} &=& \bar s (1+\gamma_5) b \, \bar s
(1+\gamma_5) s~, \nn\\
O_{\sss L}^{\sss T} &=& \bar s \sigma_{\mu \nu} (1-\gamma_5)
b \, \bar s \sigma^{\mu \nu} (1-\gamma_5) s~,\nn\\
O_{\sss R}^{\sss T} &=& \bar s \sigma_{\mu \nu} (1+\gamma_5)
b \, \bar s \sigma^{\mu \nu} (1+\gamma_5) s~.
\label{operators}
\eea
In the above operators, we take the colors of the quark
fields in each current to be the same.  This is the case in
most typical NP models (multi-Higgs-doublets, supersymmetry,
extra $Z$'s, etc.). For $S/P$ operators, a Fierz
transformation of the fermions and colors is required in
order to get a non-vanishing contribution to the production
of the final-state vector meson $\phi$ from the vacuum
(within factorization):
\bea
O_{\sss RR(LL)}^{\sss S/P} &=& -\frac{1}{2 N_c} \bar s
(1\pm \gamma_5) b \, \bar s (1\pm \gamma_5) s - \frac{1}{8
N_c} \bar s \sigma_{\mu \nu} (1\pm \gamma_5) b \, \bar s
\sigma^{\mu \nu} (1\pm \gamma_5) s~,\nn\\
O_{\sss RL(LR)}^{\sss S/P} &=& -\frac{1}{2 N_c} \bar s
\gamma_\mu (1\pm \gamma_5) b \, \bar s \gamma^\mu (1 \mp
\gamma_5) s~.
\eea
(We have neglected the octet piece coming from the color
Fierz transformation, which is justified within
factorization.)

There are also 10 operators in which one has different quark
colors in the currents. However, these can be obtained from
Eq.~(\ref{operators}) as follows.  Suppose that there is only
one type of Lorentz structure.  The effective Hamiltonian can
then be written
\bea
H_{eff} & = & B_1\bar s O_1 b \bar s O_2 s + B_2 \bar
s_{\alpha} O_1 b_{\beta} \bar s_{\beta} O_2 s_{\alpha}~,
\eea
where $B_{1,2}$ are complex coefficients, $O_{1,2}$ represent
the Lorentz structure ($S/P$, $V/A$ or $T$), and $\alpha,
\beta$ are color indices. We will call the operator
associated with $B_1$ ``color allowed'' and that associated
with $B_2$ ``color suppressed.'' In factorization, the
amplitude for $B \to \phi K$ ($K=\ks$ or $K^*$) has the
following structure:
\beq
A(B \to \phi K) = A_1+A_2 ~,
\eeq
where
\bea
A_1 & = & (B_1+B_2/N_c)\bra{K}\bar s O_1
b\ket{B}\bra{\phi}\bar s O_2 s\ket{0}, \nonumber\\
A_2 & = & (B_2+B_1/N_c)\bra{K}\bar s O_{\sss 1F}
b\ket{B}\bra{\phi}\bar s O_{\sss 2F} s\ket{0} ~.
\label{amplitude}
\eea
In the above, ${\bar s} O_{\sss 1F} b \bar s O_{\sss 2F} s$
is obtained from ${\bar s} O_1 b \bar s O_2 s$ by performing
a Fierz transformation of the fermions and the colors. The
color octet piece is neglected, because it does not lead to
the production of a $\phi$ from the vacuum.

Now, if $A_1$ or $A_2$ vanishes, or if ${\bar s} O_{\sss 1F}
b \bar s O_{\sss 2F} s$ is the same as ${\bar s} O_1 b \bar s
O_2 s$, then there is only one amplitude, and we can work
only with color-allowed operators with a general coefficient
-- the color-suppressed operators are implicitly included in
them.  As we show below, this holds for all the operators of
Eq.~(\ref{operators}).  In what follows, the key point is
that any $S/P$ operator does not contribute to the decay
because it cannot produce a $\phi$ from the vacuum. However,
$V/A$ and $T$ operators do give a nonzero contribution.
\begin{itemize}

\item Lorentz structure $(V\pm A) \times (V\pm A)$: $\bar s
O_{\sss 1F} b \bar s O_{\sss 2F} s = {\bar s} O_1 b \bar s
O_2 s$.

\item $(V\pm A) \times (V\mp A)$: Fierz transforms into an
$S/P$ operator. Thus, $A_2=0$.

\item $(S\pm P) \times (S\mp P)$: Fierz transforms into a
$V/A$ operator. Thus, $A_1=0$.

\item $(S\pm P) \times (S\pm P)$: Fierz transforms into a
combination of an $S/P$ and a $T$ operator.  Thus, $A_1=0$
and $A_2\ne 0$.

\item $T$: Fierz transforms into a combination of an $S/P$
and a $T$ operator. Thus, we effectively have $\bar s O_{\sss
1F} b \bar s O_{\sss 2F} s = \bar s O_1 b \bar s O_2 s$.

\end{itemize}
In all cases, there is only one amplitude in
Eq.~(\ref{amplitude}) above, and so the operators of
Eq.~(\ref{operators}) contain all the ``color-suppressed''
operators.

We begin by examining the case where a single NP operator is
added, contributing to the $\bsss$ amplitude.  As noted
above, this affects both $\BKstar$ and $\BKs$, and we compute
the order of magnitude of the contribution of each of the NP
operators to these decays as follows. Consider $O_{\sss
LL}^{\sss V/A}$ (for the orders of magnitude, we ignore the
$1/N_c$ coming from the inclusion of the color-suppressed
operators):
\bea
\bra{\phi K} O_{\sss LL}^{\sss V/A} \ket{\bd} &=& \bra{\phi
K} \bar s \gamma_\mu (1-\gamma_5) b \, \bar s \gamma^\mu
(1-\gamma_5) s \ket{\bd}\nn\\
&\equiv& \bra{\phi K} (V-A) \otimes (V-A) \ket{\bd}\nn\\
&=&\bra{K} V-A \ket{\bd}_\mu \bra{\phi} V-A \ket{0}^\mu\nn\\
&=&\bra{K} V \ket{\bd}_\mu \bra{\phi} V \ket{0}^\mu - \bra{K}
A \ket{\bd}_\mu \bra{\phi} V \ket{0}^\mu~.
\eea
We can now use the factorized matrix elements discussed in
appendix \ref{appendixA}. We obtain
\bea
\bra{\phi \ks} O_{\sss LL}^{\sss V/A} \ket{\bd} &=& 2 f_+
f_\phi m_{\sss B} p_c = \mathcal{O}(1),\nn\\
\left.\bra{\phi K^*} O_{\sss LL}^{\sss V/A}
\ket{\bd}\right|_{\lambda = 0} &=& -i f_\phi \frac{m_{\sss
B}+m_{\sss K^*}}{2 m_{\sss K^*}} \left\{ (m_{\sss
B}^2-m_\phi^2-m_{\sss K^*}^2)A_1 - \frac{4 m_{\sss B}^2 p_c^2
A_2 }{(m_{\sss B}+m_{\sss K^*})^2} \right\}\nn\\
&=&~ \mathcal{O}(1),\nn\\
\left.\bra{\phi K^*} O_{\sss LL}^{\sss V/A}
\ket{\bd}\right|_{\lambda = +} &=& - 2 i f_\phi V
\frac{m_\phi m_{\sss B} p_c}{m_{\sss B}+m_{\sss K^*}} + i
f_\phi m_\phi (m_{\sss B}+m_{\sss K^*})A_1\nn\\
&=& \mathcal{O}((\Lambda_{\sss QCD}/m_b)^2)~\nn\\
\left.\bra{\phi K^*} O_{\sss LL}^{\sss V/A}
\ket{\bd}\right|_{\lambda = -} &=& + 2 i f_\phi V
\frac{m_\phi m_{\sss B} p_c}{m_{\sss B}+m_{\sss K^*}} + i
f_\phi m_\phi (m_{\sss B}+m_{\sss K^*})A_1\nn\\
&=&~\mathcal{O}(\Lambda_{\sss QCD}/m_b)~,
\eea
where $p_c$ is the magnitude of the momentum of final-state
particles in the $\bd$ rest frame. The values of $f_+$,
$A_1$, $A_2$ and $V$ are given in appendix \ref{appendixA}.
Note: the polarizations in $\BKstar$ are denoted $L$
(longitudinal) and $\|$, $\perp$ (transverse). However, above
we refer to $+$ and $-$ polarizations -- the transverse
($A_{\| ,\perp }$) and helicity amplitudes ($A_\pm$) are
related by $A_{\| ,\perp }=\left(A_+ \pm A_-
\right)/\sqrt{2}$ (and $\bar A_{\| ,\perp }=\left(\bar A_-
\pm \bar A_+ \right)/\sqrt{2}$ for the CP-conjugate
amplitudes). The above $\bd \to \phi K$ amplitudes correspond
roughly to orders of magnitude $1, 1, \xi^2, \xi$, where $\xi
= \mathcal{O}(\Lambda_{\sss QCD}/m_b)$.

\begin{table}[tbh]
\center
\begin{tabular}{|c|c|c|c|c|}
\hline
& $\phi \ks$ & $\phi K^*$(L) & $\phi K^*$($-$) & $\phi
K^*$($+$) \\
\hline
$O_{\sss LL}^{\sss V/A}, O_{\sss LR}^{\sss V/A}$ & $1$ & $1$
& $\xi^2$ & $\xi$ \\
$O_{\sss RL}^{\sss V/A}, O_{\sss RR}^{\sss V/A}$ & $1$ & $1$
& $\xi$ & $\xi^2$ \\
$O_{\sss LL}^{\sss S/P}$ & $\xi^2$ & $\xi$ & $1$ & $\xi^2$
\\
$O_{\sss LR}^{\sss S/P}$ & $1$ & $1$ & $\xi^2$ & $\xi$ \\
$O_{\sss RL}^{\sss S/P}$ & $1$ & $1$ & $\xi$ & $\xi^2$ \\
$O_{\sss RR}^{\sss S/P}$ & $\xi^2$ & $\xi$ & $\xi^2$ & $1$
\\
$O_{\sss L}^{\sss T}$ & $\xi$ & $\xi$ & $1$ & $\xi^2$ \\
$O_{\sss R}^{\sss T}$ & $\xi$ & $\xi$ & $\xi^2$ & $1$ \\
\hline
\end{tabular}
\caption{Relative orders of magnitude of the contribution of
NP operators to the amplitudes of $\BKs$ and the three
polarizations of $\BKstar$ [$\xi = \mathcal{O}(\Lambda_{\sss
QCD}/m_b)$].}
\label{tab:order}
\end{table}

We have analyzed all NP operators similarly. The results are
shown in Table \ref{tab:order}. In order to generate large CP
asymmetries in $\BKs$, the contribution of the NP operator
must be large [$O(1)$]. This points to the four $O^{\sss
V/A}$ operators, $O_{\sss LR}^{\sss S/P}$ or $O_{\sss
RL}^{\sss S/P}$. In order to reproduce the $\fTfL$
measurement in $\BKstar$, the NP contribution to a transverse
polarization must be large [$O(1)$], while not contributing
significantly to the longitudinal polarization. We see that
only NP operators of the form $O_{\sss LL}^{\sss S/P}$,
$O_{\sss RR}^{\sss S/P}$, $O_{\sss L}^{\sss T}$ or $O_{\sss
R}^{\sss T}$ satisfy this criterion \cite{BphiK*NP}.

However, the key point here is that there is no NP operator
that significantly affects both $\BKs$ and a transverse
amplitude of $\BKstar$.  We therefore conclude that, as long
as $\BKs$ exhibits large CP-violating effects, there is no
single NP operator which can account for the observations in
both $\BKs$ and $\BKstar$ decays. Note: if one assumes that
there are no NP signals in this decay, which might be
justified with the present errors, then 1-NP-operator
solutions are still possible. But if one assumes that there
{\bf are} NP signals here, as would clearly be indicated if
the errors are reduced by a factor of 2, then 1-NP-operator
solutions are not possible.

We now turn to the case where two NP operators are added.
Here it is obvious that one of the operators must contribute
significantly to $\BKs$ (6 possibilities), and the other to a
transverse amplitude of $\BKstar$ (4 possibilities).  Thus,
we must in principle consider 24 pairs of operators.
However, this number can be reduced as follows.  In any
reasonable NP model, if there are only two new operators,
these are typically both of the $V/A$, $S/P$ or $T$ variety.
There are no pairs of $V/A$ or $T$ operators which give large
contributions to both decays, so there are only 4 pairs of NP
operators to examine: ($O_{\sss LL}^{\sss S/P}$, $O_{\sss
LR}^{\sss S/P}$), ($O_{\sss LL}^{\sss S/P}$, $O_{\sss
RL}^{\sss S/P}$), ($O_{\sss RR}^{\sss S/P}$, $O_{\sss
LR}^{\sss S/P}$) and ($O_{\sss RR}^{\sss S/P}$, $O_{\sss
RL}^{\sss S/P}$).  The results of Table \ref{tab:order} give
only the general size of contributions, so it is necessary to
perform a fit to see which pairs of NP operators can account
for the observations in both decays, and to what extent.  We
do this below.

\begin{table}[tbh]
\center
\begin{tabular}{|c|c|}
\hline
BR & $9.8^{+0.7} _{-0.6}\times 10^{-6}$ \\
$A_{\sss CP}$ & $0.01 \pm 0.05$ \\
$f_{\sss L}$ & $0.480 \pm 0.030$ \\
$f_\bot$ & $0.241 \pm 0.029$ \\
\hline
\end{tabular}
\caption{Measurements of $B\to \phi K^{*0}$. Included are the
  branching fraction (BR), the direct CP symmetry ($A_{\sss
  CP}$), and the fraction of longitudinal and $\bot$ decays,
  $f_{\sss L}$ and $f_\bot$ \cite{pdg,HFAG,phiK*meas}.}
\label{tab:dataVV}
\end{table}

The fit includes the three observables of $\BKs$ shown in
Table \ref{tab:dataPV}: BR, $S_{\sss CP}$, $A_{\sss CP}$. It
also includes four observables of $\BKstar$: BR, $A_{\sss
CP}$, $f_{\sss L}$, $f_\bot$. The latest values are given in
Table \ref{tab:dataVV}\footnote{In Table \ref{tab:dataVV} it
is given that $f_{\sss L} = 0.48$, $f_\bot = 0.24$.  The
$f_i$ are defined such that $f_\| = 1 - f_{\sss L} - f_\bot =
0.28$, so that indeed $\fTfL = (f_\bot + f_\|)/\fL \simeq
1$.}.  (There are other measurements of $\BKstar$, but they
are not used in this paper.)

The NP contributions to the above quantities are taken into
account as follows. We are considering the effect of the SM
and two NP operators (we generally refer to them as $O_1$ and
$O_2$). The strength of the NP is parametrized by unknown
complex coefficients (referred to as $C_1$ and $C_2$).  The
SM piece is calculable within QCD factorization (QCDf)
\cite{BBNS}, and we take the value of its contribution from
there. $C_1$ and $C_2$ generally each have a magnitude, a
weak phase, and a strong phase.  However, in Ref.~\cite{DL},
it was shown that the NP strong phases are negligible
compared to that of the (dominant) SM contribution. Thus, we
have only four free parameters: two NP magnitudes and two NP
weak phases of $C_1$ and $C_2$. The decay amplitudes
$A^\lambda$ for a given helicity $\lambda$ can be written in
terms of these free parameters simply by computing the matrix
elements for $\BKs$ and each polarization of $\BKstar$.  In
general, they take the form
\bea
A^\lambda &=& A_{\sss SM}^\lambda + \bra{\phi K} C_1 O_1 +
C_2 O_2 \ket{\bd}\Big|_\lambda\nn\\
&\equiv& A_{\sss SM}^\lambda + C_1 A_1^\lambda + C_2
A_2^\lambda~,
\eea
where $A_{1,2}$ are the factorized matrix elements given in
appendix \ref{appendixA}.  The CP-conjugate amplitudes $\bar
A$ are obtained by changing the sign of the unknown weak
phases in $C_1$ and $C_2$.

All observables can be expressed in terms of the amplitudes
$A^\lambda$. The branching ratio is given by
\beq
{\rm BR} = \frac{\tau_{\sss B} p_c}{8 \pi \hbar m_{\sss B}^2}
\sum_\lambda |A^\lambda|^2 ~,
\eeq
where $\tau_{\sss B}$ is the lifetime of the $\bd$
meson. This applies to both $\BKs$ and $\BKstar$. The
time-independent CP asymmetry is given by
\beq
A_{\sss CP} = -\,\frac{|A|^2-|\bar A|^2}{|A|^2+|\bar A|^2} ~.
\eeq
For $\BKs$, $A$ is the amplitude and ${\bar A}$ is its
CP-conjugate. For $\BKstar$, $|A|^2 = \sum_\lambda
|A^\lambda|^2$. The time-dependent CP-asymmetry in $\BKs$ is
given by
\beq
S_{\sss CP} = -2 \, \frac{\hbox{Im} (e^{-2 i \beta} A^* \bar
A)}{|A|^2+|\bar A|^2}~.
\eeq
Finally, the helicity fractions of $\BKstar$ are defined
as usual by
\beq
f_{\sss L,||,\bot}^{\sss B}= \frac{|A^{\sss
L,||,\bot}|^2}{|A^{\sss L}|^2+|A^{||}|^2+|A^\bot|^2}~,
\eeq
and similarly for the CP-conjugates $f_{\sss L,||,\bot}^{\sss
  \bar B}$.  Combining them, $f_{\sss L,||,\bot}$ are defined
  by
\beq
f_{\sss L,||,\bot} = \frac{1}{2}(f_{\sss L,||,\bot}^{\sss B}
+ f_{\sss L,||,\bot}^{\sss \bar B})~.
\eeq

The 7 observables can therefore be expressed in terms of the
4 free parameters of $C_1$ and $C_2$.  Thus, it is possible
to perform a fit to obtain the preferred values of these
parameters, and to determine whether it is possible to
account for the data with the addition of certain NP
operators.  However, there is a complication in all of this.
In the SM, the decays $\BKs$ and $\BKstar$ are dominated by
the $\btos$ penguin amplitude, $P'$. $P'$ is actually
composed of three pieces, $P'_u$, $P'_c$ and $P'_t$, where
the subscript refers to the internal quark in the loop (the
pieces $P'_{u,c}$ are rescattering amplitudes generated
mainly from tree-level operators). We can write
\bea
P' & = & V_{ub}^* V_{us} \, P'_u + V_{cb}^* V_{cs} \, P'_c +
V_{tb}^* V_{ts} \, P'_t \nn\\
&\simeq & V_{cb}^* V_{cs} \, (P'_c - P'_t) ~.
\eea
In writing the second line, we have used the unitarity of the
Cabibbo-Kobayashi-Maskawa matrix to eliminate the $V_{tb}^*
V_{ts}$ term, and we have dropped the $V_{ub}^* V_{us}$ term
since $|V_{ub}^* V_{us}| \ll |V_{cb}^* V_{cs}|$.  As the weak
phase of $V_{cb}^* V_{cs}$ is zero, $P'$ has only its
magnitude and a strong phase.  It is these quantities that
are calculated in QCDf. Unfortunately, as we see below, the
QCDf results are not very precise.

The QCDf calculation is treated by applying Refs.~\cite{BBNS,
BN, BRY} straightforwardly.  For $\BKs$,
\beq
\frac{P'_c-P'_t}{A_{\sss \phi \ks}} = \alpha_3^c + \alpha_4^c
- \frac{1}{2} \alpha_{\sss 3,EW}^c - \frac{1}{2} \alpha_{\sss
4,EW}^c \equiv \mathcal{P}_{\sss \phi \ks}~,
\eeq
where the $A_{\sss \phi \ks}$ are form factors as defined in
appendix \ref{appendixA}, and the $\alpha$'s are defined in
Ref.~\cite{BN}.  (Note: the above formula could contain
$\beta$ terms.  However, we have neglected all such pieces,
consistent with our assumption that PA is not present.) For
$\BKstar$, the SM penguin has the same form, but with
explicit polarization dependence $\lambda$:
\beq
\frac{P^{'\lambda}_c-P^{'\lambda}_t} {A_{\sss \phi
K^*}^\lambda} = \alpha_3^{c,\lambda} + \alpha_4^{c,\lambda} -
\frac{1}{2} \alpha_{\sss 3,EW}^{c,\lambda} - \frac{1}{2}
\alpha_{\sss 4,EW}^{c,\lambda} \equiv \mathcal{P}_{\sss \phi
K^*}^\lambda~.
\eeq
The $\alpha^\lambda$'s are defined in Ref.~\cite{BRY}.
Several inputs are required in order to get magnitudes and
strong phases. For quark masses, BBNS values \cite{BBNS} were
used, allowing them to vary within a range of $1\sigma$.  For
meson masses, Particle Data Group values \cite{pdg} were
used, fixed at their central value.  Wilson coefficients were
calculated using Refs.~\cite{BBL, BJL}, with the
renormalization scale $\mu$ allowed to vary within $[m_b/2 ,
m_b]$.  For decay constants, values from Table
\ref{tab:decayconstants} in appendix \ref{appendixA} were
used, within a range of $1\sigma$.  For form factors, fixed
values of Table \ref{tab:formfactor} in appendix
\ref{appendixA} were used, but we studied each of the three
cases (minimal, central and maximal values).  The allowed
ranges of the SM penguin amplitudes are summarized in Table
\ref{tab:QCDf}.  The SM penguin of negative helicity is
neglected because of the small form factors
($\mathcal{P}_{\phi K^*}^- = \bar \mathcal{P}_{\phi K^*}^+
\simeq 0$). As expected, form factors have little impact on
the values of the SM penguin amplitudes $\mathcal{P}_{\sss
\phi \ks}$ and $\mathcal{P}_{\sss \phi K^*}^\lambda$, since
they contribute at subleading order.  Numerical variations
are mainly due to the random scanning of the parameter space.

\begin{table}[tbh]
\center
\begin{tabular}{|c|c|c|c|}
\hline
 & Minimal & Central & Maximal \\
\hline
$|\mathcal{P}_{\phi \ks}|$ & $[0.031,0.064]$ & $[0.031,0.062]$  & $[0.031, 0.060]$ \\
Arg$(\mathcal{P}_{\phi \ks})$(rad) & $[3.2, 3.6]$ & $[3.2, 3.6]$  & $[3.2, 3.6]$ \\
$|\mathcal{P}_{\phi K^*}^0|$ & $[0.025, 0.036]$ & $[0.026,0.036]$  & $[0.027, 0.037]$ \\
Arg$(\mathcal{P}_{\phi K^*}^0)$(rad) & $[3.4, 3.6]$ & $[3.4, 3.6]$  & $[3.4, 3.6]$ \\
$|\mathcal{P}_{\phi K^*}^+|$ & $[0.031, 0.062]$ & $[0.033, 0.062]$  & $[0.033, 0.061]$ \\
Arg$(\mathcal{P}_{\phi K^*}^+)$(rad) & $[3.0, 3.4]$ & $[3.0, 3.4]$  & $[3.0, 3.4]$ \\
\hline
\end{tabular}
\caption{Allowed ranges for SM penguin magnitudes and strong
phases according to QCDf, for the three sets of form-factor
values.}
\label{tab:QCDf}
\end{table}

Ideally, in order to take into account the ranges of the QCDf
determinations, one would scan over the allowed regions of
all magnitudes and strong phases.  For each set of SM values,
the $\chi^2$ would be evaluated.  In this way, we could find
the best fit (i.e.\ smallest value of $\chi^2_{min}$) for
each of the 2-NP-operator solutions.  Unfortunately, this is
not possible, as the space of SM values is too large (e.g.\
if we take 10 SM values/region, we would require $10^6$
$\chi^2$ evaluations).  As a compromise, we have adopted the
following procedure: we fix the SM strong phases to their
central values, and scan over the SM magnitudes.  However, we
have checked what happens when we take different values for
the strong phases.  We find that the $\chi^2$ numbers can
change quite a bit, but a bad fit cannot be turned into a
good fit.

\begin{table}[tbh]
\center
\begin{tabular}{|c|c|c|c|}
\hline
Operators & Minimal & Central & Maximal \\
\hline
$(O_{\sss LL}^{\sss S/P}, O_{\sss RL}^{\sss S/P})$ &
2.6 (45.7\%) & 2.8 (42.4\%) & 3.1 (37.6\%) \\
$(O_{\sss LL}^{\sss S/P}, O_{\sss LR}^{\sss S/P})$ &
1.4 (70.6\%) & 1.3 (72.9\%) & 1.3 (72.9\%) \\
$(O_{\sss RR}^{\sss S/P}, O_{\sss RL}^{\sss S/P})$ &
1.9 (59.3\%) & 1.7 (63.7\%) & 1.6 (65.9\%) \\
$(O_{\sss RR}^{\sss S/P}, O_{\sss LR}^{\sss S/P})$ &
1.7 (63.7\%) & 1.7 (63.7\%) & 1.6 (65.9\%) \\
$(O_{\sss LL(RR)}^{\sss V/A}, O_{\sss RL(LR)}^{\sss V/A})$ &
15.7 (0.13\%) & 10.6 (1.4\%) & 7.1 (6.9\%) \\
$(O_{\sss R}^{\sss T}, O_{\sss L}^{\sss T})$ &
3.6 (30.8\%) & 3.6 (30.8\%) & 3.9 (27.2\%) \\
\hline
\end{tabular}
\caption{Best-fit results ($\chi^2_{min}$) for pairs of NP
operators, with present-day errors on $S_{\sss CP}$ and
$A_{\sss CP}$ in $\BKs$.  The calculation was done for the
three sets of form factors (minimal, central and maximal).}
\label{tab:results}
\end{table}

\begin{table}[tbh]
\center
\begin{tabular}{|c|c|c|c|}
\hline
Operators & Minimal & Central & Maximal \\
\hline
$(O_{\sss LL}^{\sss S/P}, O_{\sss RL}^{\sss S/P})$ &
6.3 (9.8\%) & 7.4 (6.0\%) & 8.6 (3.5\%) \\
$(O_{\sss LL}^{\sss S/P}, O_{\sss LR}^{\sss S/P})$ &
4.3 (23.1\%) & 4.0 (26.1\%) & 3.9 (27.2\%) \\
$(O_{\sss RR}^{\sss S/P}, O_{\sss RL}^{\sss S/P})$ &
5.2 (15.8\%) & 5.8 (12.2\%) & 5.6 (13.3\%) \\
$(O_{\sss RR}^{\sss S/P}, O_{\sss LR}^{\sss S/P})$ &
4.9 (17.9\%) & 4.7 (19.5\%) & 4.5 (21.2\%) \\
$(O_{\sss LL(RR)}^{\sss V/A}, O_{\sss RL(LR)}^{\sss V/A})$ &
20.3 (0.01\%) & 15.9 (0.12\%) & 10.9 (1.2\%) \\
$(O_{\sss R}^{\sss T}, O_{\sss L}^{\sss T})$ &
13.7 (0.33\%) & 13.5 (0.37\%) & 14.0 (0.29\%) \\
\hline
\end{tabular}
\caption{Best-fit results ($\chi^2_{min}$) for pairs of NP
operators, in the scenario in which the errors on $S_{\sss
CP}$ and $A_{\sss CP}$ in $\BKs$ are reduced by a factor of
2. The calculation was done for the three sets of form
factors (minimal, central and maximal).}
\label{tab:results2}
\end{table}

The results are shown in Tables \ref{tab:results} (current
errors on $S_{\sss CP}$ and $A_{\sss CP}$) and
\ref{tab:results2} (errors on $S_{\sss CP}$ and $A_{\sss CP}$
reduced by a factor of 2). Here we present the smallest value
of $\chi^2_{min}$ (best fit) for each of the 2-NP-operator
solutions. (The number in parentheses indicates the quality
of the fit, and depends on $\chi^2_{min}$ and $d.o.f.$
individually. 50\% or more is a good fit; fits which are
substantially less than 50\% are poorer.) In all cases, the
worst fit is given by a large value of $\chi^2_{min}$, with a
0\% quality of fit.

From Table \ref{tab:results}, we see that, with current
errors on $S_{\sss CP}$ and $A_{\sss CP}$, all four $S/P$
2-NP-operator solutions give good fits to the $\BKs$ and
$\BKstar$ data.  We also show the fit results for $V/A$ and
$T$ 2-NP-operator solutions.  We see that the $V/A$ solution
gives a very poor fit, but the $T$ solution, while not as
good as any of the $S/P$ pairs, is still acceptable.

If the errors on $S_{\sss CP}$ and $A_{\sss CP}$ are reduced
by a factor of 2 (Table \ref{tab:results2}), we find that no
$S/P$ 2-NP-operator hypothesis is an excellent fit to the
data.  On the other hand, none of them is ruled out, either.
The most that one can say is that $(O_{\sss LL}^{\sss S/P},
O_{\sss RL}^{\sss S/P})$ and $(O_{\sss RR}^{\sss S/P},
O_{\sss RL}^{\sss S/P})$ are disfavored, but even this is not
very strong.  On the other hand, in this case both $V/A$ and
$T$ 2-NP-operator solutions are essentially ruled out.

In both error scenarios, it is the large direct CP-asymmetry
measurement in $\BKs$ which is hardest to accommodate.  It
will be important to pay attention to this observable in the
future to determine which NP solutions are viable.

The fact that the best fit and worst fit have substantially
different $\chi^2_{min}$ shows that the contribution from the
SM is significant, and that all $\chi^2$ ranges would be
reduced quite a bit with an improved determination of the SM
values.  In fact, one could easily obtain poor fits for all
pairs of NP operators (as well as the SM).

Above, we have shown that all 2-NP-operator solutions
involving $S/P$ operators are viable, but those which contain
only $V/A$ or $T$ operators are disfavored or ruled out.
Obviously, any realistic NP model which contains more than 2
operators -- and most do -- will also be allowed, as long as
the observations in both $\BKs$ and $\BKstar$ decays are
explained. In Table \ref{tab:models}, we show which types of
operators are present for some simple NP models.  Even though
models with supersymmetry\footnote{Models with supersymmetry
generate the $\bsss$ transition mainly through squark-gluino
loops.}  or extra $Z'$ bosons typically generate several
operators, they are all of $V/A$ type.  As such, they cannot
explain both $\BKs$ and $\BKstar$ data.  On the other hand,
the two-Higgs-doublet model is favored because it contains
only $S/P$ operators (perhaps all 4 pairs), and can
potentially accomodate both $\BKs$ and $\BKstar$.

\begin{table}[tbh]
\center
\begin{tabular}{|c|c|c|c|}
\hline
Models & $V/A$ & $S/P$ & $T$ \\
\hline
Supersymmetry \cite{GNK, ACH, BBMR} & $\times$ & & \\
Two Higgs doublets \cite{2HDM1,2HDM2} &  & $\times$ &  \\
Extra $Z'$ bosons \cite{LP} & $\times$ &  &  \\
\hline
\end{tabular}
\caption{Summary of operator content for some simple NP
models.}
\label{tab:models}
\end{table}

Even with the assumption of reduced errors on the
CP-violating observables in $\BKs$, all four $S/P$
2-NP-operator solutions are allowed.  This raises the obvious
question: is there any way to distinguish these solutions?
Clearly, smaller errors on the experimental measurements
and/or the theoretical determination of the SM contribution
can help.  With these, it may be that the $\chi^2_{min}$ of
one solution is strongly preferred over that of the other
three.  

However, there is an additional possibility.  In any $B\to
V_1V_2$ decay, one can construct the {\it triple product}
(TP). In the rest frame of the $B$, the TP takes the form
${\vec q} \cdot ({\vec\varepsilon}_1 \times
{\vec\varepsilon}_2)$, where ${\vec q}$ is the momentum of
one of the final vector mesons, and ${\vec\varepsilon}_1$ and
${\vec\varepsilon}_2$ are the polarizations of $V_1$ and
$V_2$.  There are two TP's, which can be written \cite{DLTP}
\beq
A_{\sss T}^{(1)} \equiv \frac{\hbox{Im}(A_\bot
A_0^*)}{|A_0|^2+|A_{||}|^2+|A_\bot|^2}~,~~~~~A_{\sss T}^{(2)}
\equiv \frac{\hbox{Im}(A_\bot
A_{||}^*)}{|A_0|^2+|A_{||}|^2+|A_\bot|^2}~.
\eeq
The $\bar A_{\sss T}^{(1,2)}$ for ${\bar B}$ decays are
defined similarly. The TP asymmetry is defined
by\footnote{Note: in contrast to Ref.~\cite{DLTP}, this
definition involves a subtraction rather than an addition.
This is because we have defined $A_{\perp, ||}$ and $\bar
A_{\perp, ||}$ in such a way that $\mathcal{A}_{\sss
T}^{(i)}$ is zero in the absence of CP violation.}
\beq
\mathcal{A}_{\sss T}^{(i)} = \frac{A_{\sss T}^{(i)}-\bar
A_{\sss T}^{(i)}}{2}~.
\eeq
The ``fake'' TP asymmetry $\tilde \mathcal{A}_{\sss T}^{(i)}$
is given by the same definition, but the TP's are added
rather than subtracted. Note that the fake TP asymmetry can
be nonzero even if CP is conserved.

\begin{table}[tbh]
\center
\begin{tabular}{|c|c|c|c|c|}
\hline
Operators & $\mathcal{A}_{\sss T}^{(1)}$ & $\tilde
  \mathcal{A}_{\sss T}^{(1)}$ & $\mathcal{A}_{\sss T}^{(2)}$
  & $\tilde \mathcal{A}_{\sss T}^{(2)}$ \\
\hline
$(O_{\sss LL}^{\sss S/P}, O_{\sss RL}^{\sss S/P})$ &
    [$-$0.30, $-$0.27] & [0.030, 0.062] & [0.16, 0.22] &
    [$-$0.006, $-$0.004] \\
$(O_{\sss LL}^{\sss S/P}, O_{\sss LR}^{\sss S/P})$ & [0.29,
0.32] & [$-$0.008, 0.014] & [$-$0.17, $-$0.14] &
[$-$0.003, 0.000] \\
$(O_{\sss RR}^{\sss S/P}, O_{\sss RL}^{\sss S/P})$ & [0.26,
  0.28] & [$-$0.099, 0.056] & [$-$0.037, 0.090] & [$-$0.004,
  0.001] \\
$(O_{\sss RR}^{\sss S/P}, O_{\sss LR}^{\sss S/P})$ &
[$-$0.33, $-$0.31] & [$-$0.036, $-$0.011] & [$-$0.001, 0.000]
& 0.000 \\
\hline
\end{tabular}
\caption{Predictions of all $S/P$ 2-NP-operator solutions of
Table \ref{tab:results} for the central values of the real
and fake TP asymmetries in $\BKstar$. The ranges of
TP-asymmetry predictions correspond to the full variation of
form-factor values.}
\label{tab:TP}
\end{table}

The above applies to $\BKstar$.  In Table \ref{tab:TP} we
present the central values of $\mathcal{A}_{\sss T}^{(1,2)}$
and $\tilde \mathcal{A}_{\sss T}^{(1,2)}$, calculated for
each of the $S/P$ solutions shown in Table \ref{tab:results}.
The ranges correspond to the sets of form-factor values
varying from minimal to maximal.  We do not include errors
because they are very large with the current data.  In any
case, our point in presenting the results of Table
\ref{tab:TP} is the following. It is clear that different
$S/P$ 2-NP-operator solutions lead to very different patterns
of central values of predictions for the TP asymmetries.
This emphasizes the usefulness of TP's for distinguishing the
various NP solutions, and we strongly encourage
experimentalists to make such measurements.

In summary, a ``polarization puzzle'' has been observed in
$\BKstar$, namely that the fraction of transversely-polarized
decays is about equal to that of longitudinally-polarized
decays, in contrast to the expectations of the naive standard
model (SM).  In this paper, we explore the new-physics (NP)
solution to this puzzle.  We first note that any NP
explanation must also be consistent with the observations in
$\BKs$.  This decay is particularly intriguing since the
present measurements of CP-violating asymmetries are in
disagreement with the SM.  On the other hand, the errors are
still sufficiently large that this discrepancy is not
statistically significant.  As such, any constraints on NP in
$\BKstar$ are not that stringent.  For this reason, we also
perform the analysis with the assumption that future
measurements will show a greater statistical discrepancy.
That is, we use the central values of the CP-asymmetry
measurements, but take the errors to be reduced by a factor
of 2.

We consider 10 NP operators, of types $S/P$, $V/A$ and $T$.
We first show that, as long as $\BKs$ exhibits large
CP-violating effects, no single NP operator can explain the
data in both $\BKstar$ and $\BKs$ decays.  Turning to
2-NP-operator solutions, it is clear that one of the
operators must contribute significantly to $\BKs$, and the
other to a transverse amplitude of $\BKstar$.  In any
realistic NP model the two operators are typically both of
the $V/A$, $S/P$ or $T$ type.  However, no pairs of $V/A$ or
$T$ operators give large contributions to both decays, so
that only 4 pairs of $S/P$ operators need be considered.  We
have performed fits to several observables in $\BKstar$ and
$\BKs$ decays, and find that all four 2-NP-operator solutions
are allowed.  (We also show explicitly that the $V/A$ and $T$
solutions are disfavored or ruled out.)

One can distinguish among the solutions in several ways.  If
the experimental errors on future measurements are improved,
one solution might be preferred.  Alternatively, the
theoretical uncertainty can be reduced if the SM contribution
to $\BKstar$ and $\BKs$ is better determined.  Finally, the
four solutions predict a very different pattern of
triple-product asymmetries in $\BKstar$.  Their measurement
could help distinguish among the possible NP solutions.

Finally, any realistic NP model which contains more than two
operators will also be allowed, as long as the measurements
in both $\BKs$ and $\BKstar$ decays are explained.  However,
models with supersymmetry or extra $Z'$ bosons contain only
operators of $V/A$ type, and therefore cannot explain both
$\BKs$ and $\BKstar$ data.  In contrast, the
two-Higgs-doublet model has only $S/P$ operators, and is thus
favored.

\bigskip
\noindent {\bf Acknowledgements}:
We thank J. Matias and M. Nagashima for helpful
communications. This work is financially supported by NSERC
of Canada.


\appendix

\section{Calculation of factorized matrix elements}
\label{appendixA}

For NP, we work within the framework of naive factorization.
Since LO contributions of NP are expected to be subleading
compared with the SM, NLO contributions of NP can be safely
neglected. Then, for a general effective four-quark operator
$O\sim X\otimes Y$ ($X,Y=S, P, V, A, T$ or $T \gamma_5$), the
matrix element is assumed to factorize as
\beq
\bra{\phi K} O \ket{B} \to \bra{K} X \ket{B} \bra{\phi} Y
\ket{0}~,
\eeq
where $K$ stands for $\ks$ or $K^*$.  $\bra{K} X \ket{B}$ is
calculable using known form factors; $\bra{\phi} Y \ket{0}$
is calculable using the $\phi$-meson decay constants.

For $B \to K$ form factors, we use definitions from
Refs.~\cite{ABHH, BJZ}:
\bea
\bra{K(p)}V\ket{\bar B(p_{\sss B})}_\mu &=& f_+(s) \left\{
(p_{\sss B}+p)_\mu - \frac{m_{\sss B}^2-m_K^2}{s} q_\mu
\right\} \nn\\
&& ~~~~~~~~~~+ \frac{m_{\sss B}^2-m_K^2}{s} f_0(s)
q_\mu~,\nn\\
\bra{K(p)}T(1\pm \gamma_5) \ket{\bar B(p_{\sss B})}_{\mu \nu}
q^\nu&=& \bra{K(p)} T \ket{\bar B(p_{\sss B})}_{\mu \nu}
q^\nu \nn\\
&=&i \left\{ (p_{\sss B}+p)_\mu s - q_\mu (m_{\sss B}^2
-m_K^2) \right\} \frac{f_{\sss T}(s)}{m_{\sss B}+m_K}~,\nn\\
\bra{K^*(p,\epsilon)} V\pm A \ket{\bar B(p_{\sss B})}_\mu &=&
\pm i \epsilon_\mu^* (m_{\sss B}+m_{\sss K^*}) A_1(s) \nn\\
& & \mp i (p_{\sss B}+p)_\mu (\epsilon^* \cdot p_{\sss B})
\frac{A_2(s)}{m_{\sss B}+m_{\sss K^*}}\nn\\
&&\hskip-3truecm \mp i q_\mu (\epsilon^* \cdot p_{\sss
B}) \frac{2 m_{\sss K^*}}{s} (A_3(s)-A_0(s))+\epsilon_{\mu
\nu \rho \sigma} \epsilon^{*\nu}p_{\sss B}^\rho p^\sigma
\frac{2 V(s)}{m_{\sss B}+m_{\sss K^*}}\nn\\
\bra{K^*(p, \epsilon)} T (1\pm \gamma_5) \ket{\bar B(p_{\sss
B})}_{\mu \nu}q^\nu &=& i \epsilon_{\mu \nu \rho \sigma}
\epsilon^{*\nu} p_{\sss B}^\rho p^\sigma 2 T_1(s)\nn\\
&&\hskip-1truecm \pm T_2(s)\left\{ \epsilon^*_\mu (m_{\sss
B}^2-m_{\sss K^*}^2)-(\epsilon^*\cdot p_{\sss B})(p_{\sss
B}+p)_\mu \right\}\nn\\
&&\hskip-1truecm \pm T_3(s)(\epsilon^* \cdot p_{\sss B})
\left\{ q_\mu - \frac{s}{m_{\sss B}^2-m_{\sss K^*}^2}
(p_{\sss B}+p)_\mu \right\}~,
\eea
with $q=p_{\sss B}-p$ and $s=q^2$.  Values of the form
factors are tabulated in Tables 3, 4 and 5 of
Ref.~\cite{ABHH} with minimal and maximal values and $s$
dependance.  Using these, we have calculated them for the
case of $s = m_\phi^2$ (see Table \ref{tab:formfactor}).

\begin{table}[tbh]
\center
\begin{tabular}{cccc}
\hline
\hline
& Minimal value & Central value & Maximal value \\
\hline
$f_+$ & $0.295$ & $0.337$ & $0.391$ \\
$f_0$ & $0.286$ & $0.327$ & $0.379$ \\
$f_{\sss T}$ & $0.319$ & $0.375$ & $0.446$ \\
\hline
$A_1$ & $0.301$ & $0.345$ & $0.393$ \\
$A_2$ & $0.258$ & $0.295$ & $0.333$ \\
$A_0$ & $0.437$ & $0.498$ & $0.750$ \\
$V$ & $0.423$ & $0.483$ & $0.579$ \\
\hline
$T_1$ & $0.355$ & $0.402$ & $0.463$ \\
$T_2$ & $0.341$ & $0.387$ & $0.445$ \\
$T_3$ & $0.245$ & $0.272$ & $0.307$ \\
\hline
\hline
\end{tabular}
\caption{ Values of $B \to K$, $B \to K^*$ form factors for
$s=m_\phi^2$ following Ref. \cite{ABHH} (calculated in the
QCD light-cone sum-rules approach at the scale $\mu =m_b$).}
\label{tab:formfactor}
\end{table}

The $\phi$ vector-meson decay constants are defined by
\cite{ABHH, BJZ}
\bea
\bra{\phi (q, \epsilon)}V\ket{0}^\mu &=& f_\phi m_\phi
\epsilon^{*\mu}~,\nn\\
\bra{\phi (q, \epsilon)} T \ket{0}^{\mu \nu} &=& -i
f_\phi^\bot (\epsilon^{*\mu} q^\nu - \epsilon^{*\nu}
q^\mu)~,\nn\\
\eea
which imply
\beq
\bra{\phi (q, \epsilon)} T \gamma_5 \ket{0}^{\mu \nu} =
-\frac{1}{2}f_\phi^\bot \epsilon^{\mu \nu \rho \sigma}
(\epsilon^*_\rho q_\sigma - \epsilon^*_\sigma q_\rho) ~.
\eeq
Values of the decay constants are tabulated in Table
\ref{tab:decayconstants}.

\begin{table}[tbh]
\center
\begin{tabular}{ccccc}
\hline
\hline
M & $\phi$ [MeV] & $B$ [MeV] & $K$ [MeV] & $K^*$ [MeV]\\
\hline
$f_{\sss M}$ & $215\pm 5$ & $200 \pm 25$ & $160$ & $220\pm 5$ \\
$f_{\sss M}^\bot$ & $186 \pm 9$ & $-$ & $-$ & $185 \pm 10$ \\
\hline
\hline
\end{tabular}
\caption{ Values of decay constants for mesons.\cite{BJZ}}
\label{tab:decayconstants}
\end{table}

In order to calculate factorized matrix elements, we define
the four-momemta
\beq
p_{\sss B} = (m_{\sss B},0,0,0)~,~~~~~p_{\sss K,K^*} =
(E_{\sss K,K^*},0,0,-p_c)~,~~~~~p_\phi = (E_\phi,0,0,p_c)~,
\eeq
and polarization 4-vectors
\bea
\epsilon^0_{\sss K^*} = \frac{1}{m_{\sss K^*}} (p_c,
0,0,-E_{\sss K^*})~,&&~~~~~\epsilon^\pm_{\sss
K^*}=\frac{1}{\sqrt{2}}(0,\mp 1, +i, 0)~,\nn\\
\epsilon^0_{\phi} = \frac{1}{m_{\phi}} (p_c,
0,0,E_{\phi})~,&&~~~~~\epsilon^\pm_{\phi} =
\frac{1}{\sqrt{2}}(0,\mp 1, -i, 0)~,
\eea
in the $B$-meson rest frame.  From here it is straightforward
to calculate factorized matrix elements using the above.
Those which are non-zero are
\bea
\bra{K} V \ket{\bar B}_\mu \bra{\phi} V\ket{0}^\mu &=& 2 f_+
f_\phi m_{\sss B} p_c ~,\nn\\
\bra{K} T \ket{\bar B}_{\mu \nu}\bra{\phi} T \ket{0}^{\mu
\nu} &=& 4 f_{\sss T} f_\phi^\bot \frac{m_\phi m_{\sss B}
p_c}{m_{\sss B}+m_K}~,
\eea
for $\BKs$ and
\bea
\left. \bra{K^*} T \ket{\bar B}_{\mu \nu}\bra{\phi} T
\ket{0}^{\mu \nu} \right|_{\lambda=\pm} &=& \left. \bra{K^*}
T \gamma_5 \ket{\bar B}_{\mu \nu}\bra{\phi} T \gamma_5
\ket{0}^{\mu \nu}\right|_{\lambda=\pm} \nn\\
& = & \mp 4 i f_\phi^\bot T_1 m_{\sss B} p_c~,\nn\\
\left.\bra{K^*} T \ket{\bar B}_{\mu \nu}\bra{\phi} T \gamma_5
\ket{0}^{\mu \nu}\right|_{\lambda=0}&=& i
\frac{f_\phi^\bot}{m_\phi m_{\sss K^*}} \left\{ -T_2 (m_{\sss
B}^2-m_\phi^2-m_{\sss K^*}^2) (m_{\sss B}^2-m_{\sss K^*}^2)
\right.\nn\\
&&\left.+4 p_c^2 m_{\sss B}^2\left[ T_2+T_3 m_\phi^2/(m_{\sss
B}^2-m_{\sss K^*}^2) \right] \right\}~,\nn\\
\left.\bra{K^*} T \ket{\bar B}_{\mu \nu}\bra{\phi} T \gamma_5
\ket{0}^{\mu \nu}\right|_{\lambda=\pm} &=& 2 i f_\phi^\bot
T_2 (m_{\sss B}^2-m_{\sss K^*}^2) ~,\nn\\
\left.\bra{K^*} V \ket{\bar B}_\mu \bra{\phi} V\ket{0}^\mu
\right|_{\lambda=\pm} &=& \mp 2 i f_\phi V \frac{m_\phi
m_{\sss B} p_c}{m_{\sss B}+m_{\sss K^*}} ~,\nn\\
\left.\bra{K^*} A \ket{\bar B}_\mu \bra{\phi} V\ket{0}^\mu
\right|_{\lambda=0}&=&i f_\phi \frac{m_{\sss B}+m_{\sss
K^*}}{2 m_{\sss K^*}} \left\{ (m_{\sss B}^2-m_\phi^2-m_{\sss
K^*}^2)A_1 \right. \nn\\
&& ~~~~~~~~~~~~~~~~~~~~~ \left. - \frac{4 m_{\sss B}^2 p_c^2
A_2 }{(m_{\sss B}+m_{\sss K^*})^2} \right\}~,\nn\\
\left.\bra{K^*} A \ket{\bar B}_\mu \bra{\phi} V\ket{0}^\mu
\right|_{\lambda=\pm}&=&-i f_\phi m_\phi (m_{\sss B}+m_{\sss
K^*})A_1
\eea
for $\BKstar$.  All of this allows us to calculate the
entries in Table \ref{tab:order}.

\end{document}